\documentclass[sigconf,nonacm]{acmart}

\usepackage{multirow}

\newlength{\figcol}
\newlength{\galw}

\setcopyright{none}
\renewcommand\footnotetextcopyrightpermission[1]{}

\begin{document}
\title{Explicit LOD-Conditioned Anisotropic Filtering for Neural Texture Compression}

\author{Shin Fujieda}
\affiliation{%
  \institution{Advanced Micro Devices, Inc.}
  \city{Tokyo}
  \country{Japan}}
\email{Shin.Fujieda@amd.com}
\orcid{0000-0002-2472-7365}

\author{Takahiro Harada}
\affiliation{%
  \institution{Advanced Micro Devices, Inc.}
  \city{Santa Clara}
  \country{USA}}
\email{Takahiro.Harada@amd.com}
\orcid{0000-0001-5158-8455}

\renewcommand\shortauthors{Shin Fujieda and Takahiro Harada}

\begin{abstract}
Neural texture compression (NTC) stores material textures compactly, but the dominant feature-pyramid design receives no filtering from the GPU texture unit:
it learns each material at a fixed UV and level of detail and has no explicit anisotropic-filtering (AF) mechanism.
Setting the filter from the footprint aspect ratio alone is not enough: a single input cannot carry both the shape of the footprint and its size, so the image is too blurry near the camera by the same amount at every distance.
Our method, AF-NTC, instead conditions the decoder on the same three bounded values a hardware sampler uses: minor-axis scale (LOD), aspect ratio, and major-axis orientation.
These drive a tap line along the major axis and a level blend across the minor axis, and we train against anisotropically filtered ground truth.
At $25\%$ of a full BC7 mip chain, AF-NTC beats the no-AF trilinear-BC7 baseline on all ten material--scene pairs we compare ($+1.3$ to $+7.6$\,dB), and it is $0.0$ to $3.4$\,dB below the $4\times$-larger anisotropic-BC7, depending on the material.
\end{abstract}

%
%


%
%


\begin{teaserfigure}
    \centering
    \setlength{\tabcolsep}{1.5pt}
    \begin{tabular}{@{}cccc@{}}
        {\scriptsize Naive (ratio-only)} & {\scriptsize trilinear-BC7} &
        {\scriptsize AF-NTC} & {\scriptsize Reference (supersampled)} \\
        \includegraphics[width=\figcol]{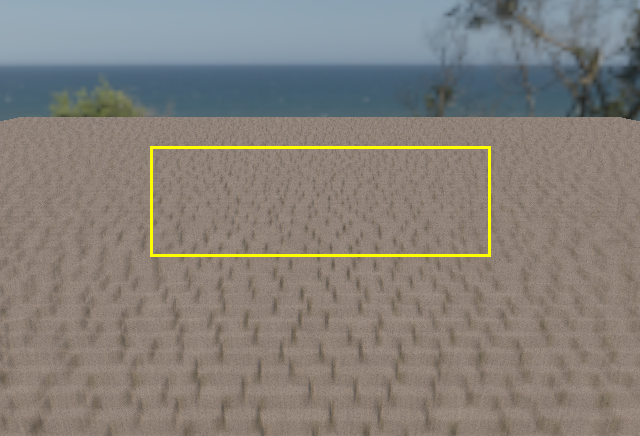} &
        \includegraphics[width=\figcol]{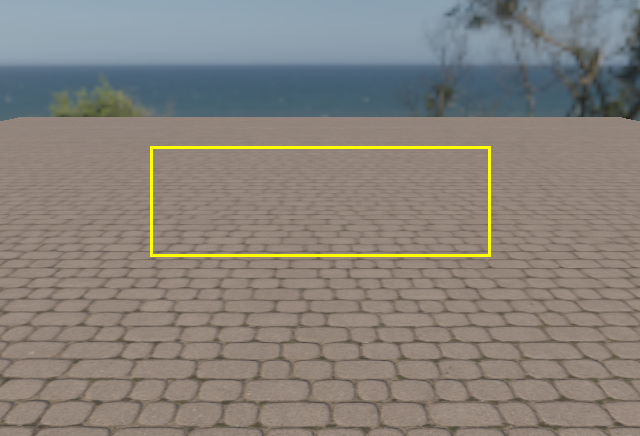}   &
        \includegraphics[width=\figcol]{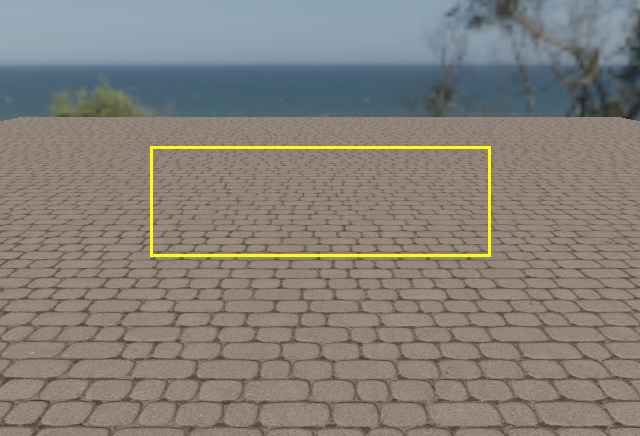}   &
        \includegraphics[width=\figcol]{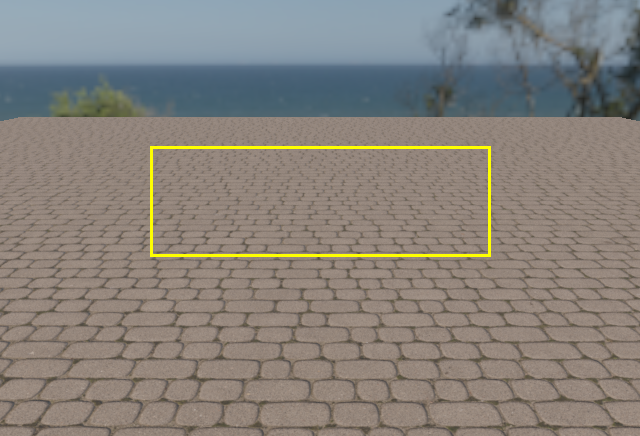}   \\[1pt]
        \includegraphics[width=\figcol]{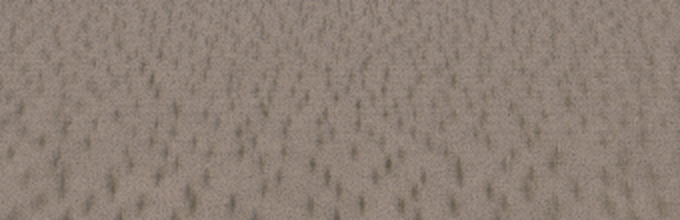} &
        \includegraphics[width=\figcol]{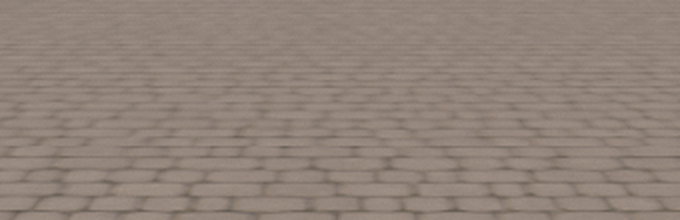}   &
        \includegraphics[width=\figcol]{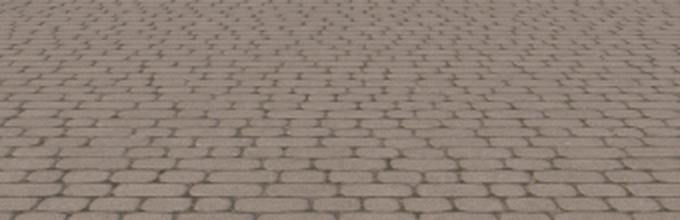}   &
        \includegraphics[width=\figcol]{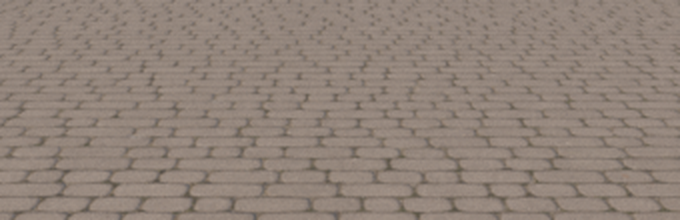}   \\
    \end{tabular}
    \caption{
        AF-NTC adds explicit anisotropic filtering to a feature-pyramid decoder at $25\%$ of the BC7 budget (\texttt{pavement} plane; render vs.\ a supersampled-EWA reference; bottom rows zooms the far parts).
        Our \emph{naive} decoder sets the filter from the aspect ratio alone, so it is too blurry even near the camera.
        The no-AF \emph{trilinear-BC7} baseline, at $4\times$ our size, is too blurry far away.
        \emph{AF-NTC} is sharp near the camera and blurry far away, and is the closest of the three to the \emph{reference}.
    }
    \Description{
        A four-column comparison of the same pavement plane seen at a grazing angle, rendered by four methods.
        From left to right: a naive ratio-only decoder, trilinear-BC7, AF-NTC, and a supersampled reference.
        The top row shows the whole frame, and the bottom row shows a zoom on the far part of the plane.
        The naive result is blurred by the same amount everywhere, including close to the camera.
        Trilinear-BC7 is sharp close to the camera but loses the pavement pattern in the distance.
        AF-NTC keeps the near detail and still reproduces the far pattern, and of the three, it looks closest to the reference.
    }
    \label{fig:teaser}
\end{teaserfigure}

\maketitle

\section{Introduction}

Neural texture compression (NTC) aims to store material textures below the on-disk size of fixed-rate block compression (BC1--BC7 depending on texture types), at equal or better quality~\cite{vaidyanathan2023ntc,bcn}.
A texture is sampled many times during rendering under minification and at grazing angles, so anisotropic filtering (AF) is necessary: it keeps detail near the camera and correctly removes it with distance.
NTC must therefore be evaluated not only on compressed size but also on filtering quality.

Two families of NTC handle AF in opposite ways.
\emph{Latent-in-BC} methods store learned features inside BC textures and let the GPU texture unit filter the latents in hardware, so AF is provided by the hardware sampler at no additional cost~\cite{belcour2025coopvec,weinreich2024bcf}.
\emph{Feature-pyramid} NTC instead learns multi-resolution feature grids with a small multilayer perceptron (MLP) decoder~\cite{vaidyanathan2023ntc}.
It has no hardware filtering and no explicit anisotropic filtering mechanism, so it aliases at one sample per pixel and relies on stochastic texture filtering (STF)~\cite{stf2024}.
This work makes AF an \emph{explicit} operation in a feature-pyramid decoder, conditioned on the pixel footprint and trained against anisotropically filtered ground truth, so that the result is sharp near the camera and blurry far away, which a naive tap blend cannot do.
We call this decoder AF-NTC.

Because the aspect ratio alone cannot carry the footprint's size, we instead condition the decoder on the same bounded $(\texttt{rot}$, $\texttt{stretch}$, $\textrm{LOD})$ triple that hardware AF uses (i.e., minor-axis scale (level of detail (LOD)), aspect ratio (\texttt{stretch}), and major-axis orientation (\texttt{rot})), and drive it along two paths: a major-axis tap line and a minor-axis level blend.
On the \texttt{pavement} scene in Fig.~\ref{fig:teaser}, AF-NTC is sharp near the camera and blurry far away, and stays closer to a supersampled reference than the no-AF baseline.

Our contributions are:
\begin{description}
  \item[Diagnosis and a minimal fix.] The aspect ratio alone cannot set the filter; the bounded $(\texttt{rot},\texttt{stretch},\textrm{LOD})$ triple restores the quality.
  \item[Two-axis learned filter.] For a decoder without hardware filtering, we combine a major-axis tap-line length with a minor-axis level blend.
  \item[Compression vs.\ filtering.] An ablation on model size shows which part of the error is compression and which is filtering.
\end{description}

\section{Related Work}
\label{sec:related}

\paragraph{Feature-pyramid NTC}
The dominant NTC design stores a material as a stack of learnable multi-resolution feature grids plus a small MLP decoder, and reports state-of-the-art quality at a matched on-disk size~\cite{vaidyanathan2023ntc}.
NTBC~\cite{fujieda2024ntbc} follows the same idea but packs the neural features into BC blocks so that a standard asset pipeline can carry them.
These decoders receive no filtering from the GPU texture unit. They learn the material at fixed UV and LOD, so one sample per pixel aliases, and practical systems rely on STF, which samples a random subset of taps, reducing temporal aliasing at the cost of added noise~\cite{stf2024}.
\citet{vaidyanathan2023ntc} report that software AF requires too many taps for real time. We target this family.

\paragraph{Latent-in-BC NTC}
A second family stores latent feature vectors inside BC textures, lets the GPU sampler filter the \emph{latents} in hardware, and then decodes with a tiny MLP~\cite{weinreich2024bcf,belcour2025coopvec}.
Anisotropic filtering is then effectively \emph{free}: no filtering logic is added to the decoder, a property we call \emph{free AF}.
A network trained only on isotropic samples generalizes to well-filtered output, because the sampler gives the actual, bounded footprint, and AF is just a weighted blend of taps over that footprint.
This is the property a feature-pyramid decoder lacks: it sees only UV, so the footprint must be an \emph{explicit, learned} input.

\paragraph{Hardware AF, EWA, and the footprint}
Classic high-quality filtering, such as the elliptical weighted average (EWA)~\cite{ewa1989}, integrates the texture over the elliptical footprint of a pixel.
Approximated hardware AF samples several trilinear taps along the footprint's major axis over a mip pyramid~\cite{williams1983mip,mccormack1999feline}.
Hardware AF derives the footprint ellipse from the screen-to-UV Jacobian; we give the derivation and the bounded inputs the sampler uses in Sec.~\ref{sec:triple}.

\begin{figure}[t]
     \centering
     \includegraphics[width=0.9\linewidth]{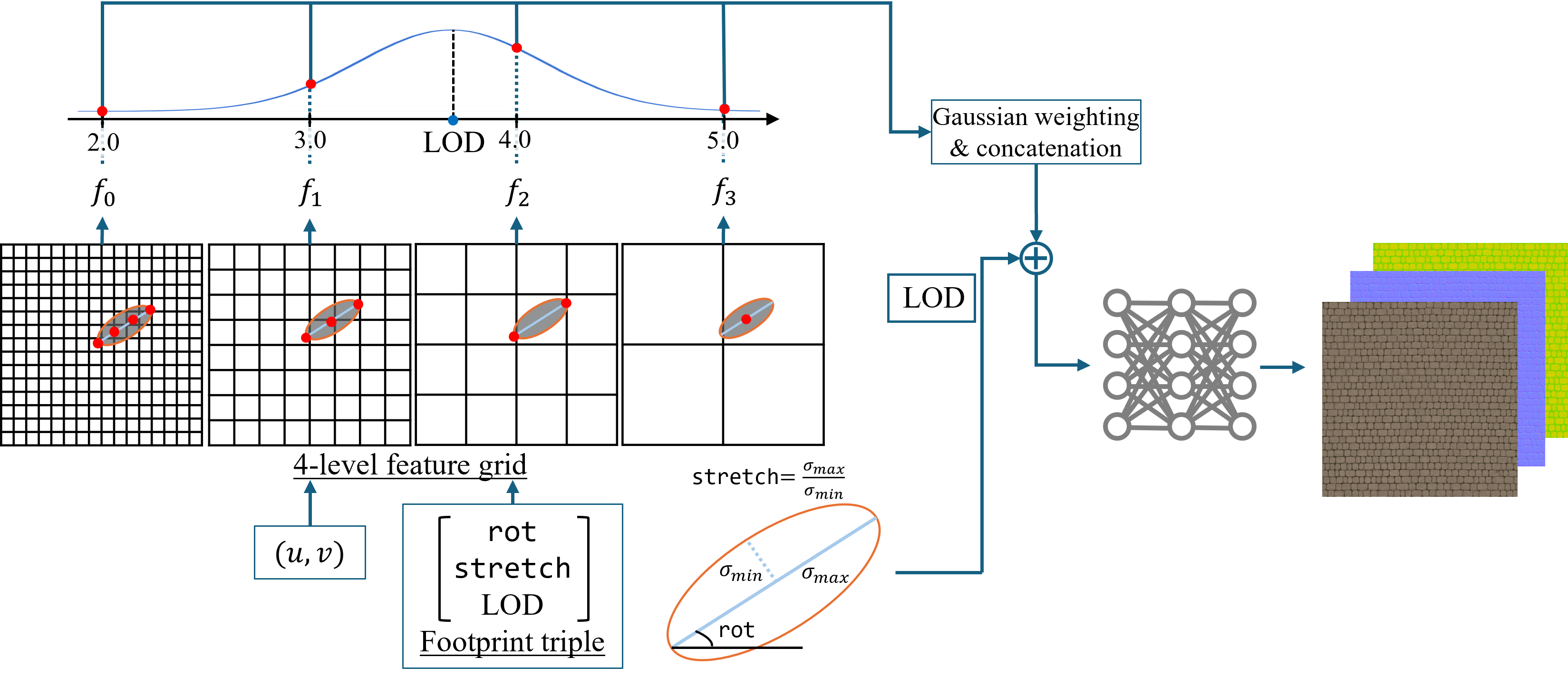}
     \caption{
          Overview of AF-NTC. The footprint triple sets the tap geometry on each grid level (bottom), the LOD-weighted blend combines levels (top), and only the concatenated feature vector and LOD reach the MLP.
     }
     \Description{
          A block diagram of the AF-NTC pipeline, read from left to right.
          At the bottom left, the UV coordinate and the footprint triple of rotation, stretch, and level of detail enter a four-level feature grid, drawn as four squares of decreasing resolution.
          On each grid, an orange ellipse marks the pixel footprint, and red dots along its major axis mark the taps; the number of taps falls as the grid gets coarser.
          At the bottom right, a larger copy of the ellipse defines stretch as the ratio of the major axis sigma-max to the minor axis sigma-min, and rot as the angle of the major axis.
          At the top, a bell-shaped curve over the level-of-detail axis weights the four levels, whose centres are 2, 3, 4 and 5; the weighted per-level features are then concatenated.
          The concatenated feature vector is joined with the level of detail and fed to a small multilayer perceptron, which outputs the decoded material maps shown as a stack of textures on the right.
     }
     \label{fig:method}
\end{figure}

\section{Method}
\label{sec:method}

AF-NTC makes AF explicit inside the feature-pyramid decoder.
Fig.~\ref{fig:method} shows the pipeline of our proposed method.

\subsection{Failure of Ratio-only Conditioning}
\label{sec:rankdef}

The naive anisotropy-aware design derives its filter length from \texttt{stretch} alone, e.g.,\ a length of the form
\begin{equation}
\texttt{length} = \texttt{stretch}\cdot k,
\label{eq:naive}
\end{equation}
with \texttt{stretch} the ratio of the long to short footprint axis and $k$ a single fixed constant.

This fails because the correct filter is set by two things that vary independently: the footprint's absolute size (how much to blur (i.e., mip level)) and its aspect ratio (how elongated it is).
\texttt{stretch} encodes only the ratio, so a ratio-only decoder must apply a fixed size $k$ that is correct at only one distance; it can never represent a near and grazing footprint, which is small yet strongly elongated.
The \textit{Naive} column of Fig.~\ref{fig:teaser} shows the result: the image is too blurry near the camera, by the same amount at every distance.
The missing quantity, absolute size, must therefore be an input of its own, which motivates the bounded $(\texttt{rot},\texttt{stretch},\textrm{LOD})$ triple.

\subsection{Bounded Hardware-AF Footprint Triple}
\label{sec:triple}

From the screen-to-UV Jacobian $J$, we form the metric $A=J\,J^{\!\top}$, whose eigenvalues are $\sigma_{\max}^2\ge\sigma_{\min}^2$;
their square roots $\sigma_{\max}\ge\sigma_{\min}$ are the major- and minor-axis lengths, and the major eigenvector gives the orientation.
From these, we form the triple as follows:
\begin{align}
\textrm{LOD} &= \log_2\!\big(\sigma_{\min}\cdot \texttt{tex\_res}\big), &&\in [0,12]\ \text{at 4K},\label{eq:lod}\\
\texttt{stretch} &= \sigma_{\max}/\sigma_{\min}, &&\in [1,16],\\
\texttt{rot} &= \angle(\text{major axis}), &&\in [0,\pi).
\end{align}
$\sigma_{\min}$ sets the LOD level, $\texttt{stretch}$ sets the tap count and, together with $\sigma_{\min}$, the tap-line length, and $\texttt{rot}$ gives the major-axis orientation.
The renderer supplies LOD per pixel; from it, we recover the minor-axis length by inverting Eq.~\ref{eq:lod}:
\begin{equation}
\sigma_{\min} = 2^{\textrm{LOD}}/\,\texttt{tex\_res}.
\label{eq:inverse}
\end{equation}
LOD and $\sigma_{\min}$ should be consistent, and we use this single recovered $\sigma_{\min}$ everywhere it is needed: both to place the taps (Eq.~\ref{eq:lenmajor}) and to build the anisotropically filtered ground truth (Sec.~\ref{sec:training}).
An elliptical footprint is fully described by its size, elongation, and orientation, and the triple $(\texttt{rot},\texttt{stretch}, \textrm{LOD})$ gives exactly these three.

\subsection{Two-Axis Learned Filter}
\label{sec:twopart}

The decoder reads input features from a four-level feature pyramid of $2$D grids, each storing a $12$d feature, bilinearly sampled on each tap and concatenated across levels into a $48$d grid-feature vector.
The footprint triple conditions the decoder both geometrically and as explicit inputs.
Geometrically, as shown in Fig.~\ref{fig:method}, it sets how the grids are filtered along the two footprint axes:
$\texttt{rot}$ orients the major axis, $\sigma_{\min}$, which is recovered from LOD with Eq.~\ref{eq:inverse}, and $\texttt{stretch}$ together set the major-axis tap-line length (Eq.~\ref{eq:lenmajor}), and LOD selects the mip level and weights that the minor-axis level blend.
As explicit inputs, the $48$d grid-feature vector is concatenated with the single normalized scalar $\textrm{LOD}/\texttt{lod\_max}$, forming the $49$d vector that feeds the MLP;
$\texttt{rot}$ and $\texttt{stretch}$ are used only through the tap geometry, as tap orientation and tap-line length, not as MLP inputs.
An ablation in Appendix~\ref{sec:supp-mlpinput} shows that also feeding $(\texttt{rot},\texttt{stretch})$ to the MLP as an $18$d positional encoding changes quality only slightly, because the tap geometry already uses $\texttt{rot}$ and $\texttt{stretch}$.

\paragraph{Major axis: absolute tap-line length}
The major-axis tap-line length is set from the \emph{absolute} minor-axis length, not the ratio:
\begin{equation}
\texttt{length\_major} = \sigma_{\min}\cdot \texttt{stretch}\quad(=\sigma_{\max}),
\label{eq:lenmajor}
\end{equation}
with $\sigma_{\min}$ from Eq.~\ref{eq:inverse}.
Unlike the naive form of Eq.~\ref{eq:naive}, where \texttt{stretch} is multiplied by the fixed constant $k$, here it is multiplied by the absolute length $\sigma_{\min}$, so the tap line spans the true major-axis length $\sigma_{\max}$ rather than a distance-independent length.
Taps are placed along the major axis, one budget per grid level, with fixed counts $[16,8,4,2]$ ($30$ taps in total) from the finest to the coarsest level.
We choose this $30$-tap budget to match hardware-AF tap counts and lower decode cost rather than for peak quality.

\paragraph{Minor axis: level blend}
Taps along the major axis alone leave the minor axis unfiltered.
We band-limit the minor axis with an LOD-weighted blend across the four grid levels.
Each level $l$ has an LOD center,
\begin{equation}
c_l=\log_2(\texttt{tex\_res}/\texttt{res}_l),
\label{eq:lodcenter}
\end{equation}
and its $12$d feature is weighted by a Gaussian in LOD about its center with $\sigma=0.7$. Then, the weighted features are concatenated, keeping the grid-feature width fixed at $48$.

\subsection{Training, Architecture, and Model Size}
\label{sec:training}

\paragraph{Ground truth}
The training target \emph{simulates} hardware AF instead of computing an exact EWA integral: we place $\lceil\texttt{stretch}\rceil$ taps along the major axis over a length of $(2^{\textrm{LOD}}/\texttt{tex\_res})\cdot\texttt{stretch}$, and fetch each tap trilinearly at $\text{mip}=\textrm{LOD}$, which filters the minor axis.
It is therefore an approximation: it agrees with a supersampled EWA reference to $37$--$41$\,dB PSNR on our scenes.
We use it only for training and measure quality on the rendered image (Sec.~\ref{sec:results}).
Note that the decoder's tap budget $[16,8,4,2]$ is fixed, while this target uses $\lceil\texttt{stretch}\rceil$ taps.

\paragraph{Filter-parameter sampling}
Instead of generating training footprints from rendered views, we sample the $(\texttt{rot},\texttt{stretch},\textrm{LOD})$ triple directly from parameter space, independently of one another.
$\texttt{rot}$ and $\texttt{stretch}$ are sampled uniformly over their ranges, and $\textrm{LOD}$ is sampled area-proportionally, $p(\textrm{LOD})\propto 4^{-\textrm{LOD}}$, so that coverage matches the screen area each LOD occupies.
Independent sampling gives the decoder footprint size and ratio as uncorrelated inputs, so it sees combinations that no single view provides.

\paragraph{Training}
As in other NTC work, our model is fitted to one 4K material and shipped with it, so training and evaluation share that texture set.
The model is implemented, trained, and evaluated in PyTorch.
We train our model with an $L_2$ loss, with optional LOD-weighting and the normal-map channels up-weighted.
We then quantize to int8 by quantization-aware training (QAT) with the Brevitas quantization library~\cite{brevitas}.

\paragraph{Architecture and model size}
The encoder holds a four-level $2$D feature grid at resolutions $1024/512/256/128$, so the per-level LOD centers $c_l$ in Eq.~\ref{eq:lodcenter} are $2,3,4,5$.
The decoder is an MLP that maps the $49$d conditioned feature vector through hidden widths $[192,192]$, leaky-ReLU activations, and a final sigmoid, outputting $9$ channels: albedo ($3$), an ARM channel packing ambient occlusion, roughness, and metallic into one texture ($3$), and a normal ($3$).
AF-NTC has an on-disk size of $16.75$\,MB (int8-packed grids and MLP), about $25\%$ of the $67$\,MB that these same three 4K maps (albedo, ARM, and normal) occupy as a full BC7 mip chain.

\section{Results}
\label{sec:results}

\paragraph{Setup}
All our numbers are PSNR measured in \emph{rendering space} rather than texture space: we render with the maps that each method produces and compare against a supersampled EWA ground truth rendered under the same lighting, so the two images differ only in the maps.
We evaluate five 4K materials, ordered from low to high spatial frequency in Table~\ref{tab:main}, each rendered in two scenes: a flat \texttt{plane} (sharp near, blurred far) and a curved \texttt{sphere}.
We compare AF-NTC against two BC7 baselines that use the same software AF implementation as AF-NTC and store the full mip chain, about $4\times$ the AF-NTC size: \emph{aniso-BC7} (full anisotropic filtering, a strong upper reference) and \emph{trilinear-BC7} (the same BC7 maps filtered only isotropically, with trilinear mip sampling and no AF, a weaker baseline).
Unless noted, we default to the int8 model (Sec.~\ref{sec:training}).

\paragraph{Conditioning}
Table~\ref{tab:main} builds up AF-NTC one part at a time on top of an \textit{isotropic} baseline: first training on anisotropic ground truth (\textit{+aniso}), then feeding LOD to the decoder (AF-NTC).
The isotropic baseline follows the free-AF approach~\cite{weinreich2024bcf,belcour2025coopvec}: the same decoder trained on isotropic samples only, with no footprint input.
Learning the AF footprint explicitly improves it throughout: each part raises PSNR on all materials and both scenes.
The gain is smallest on low-frequency \texttt{blue\_metal} ($+1.02$\,dB plane) and largest on the high-frequency materials, up to $+4.66$\,dB (plane, \texttt{stone\_floor}) and $+5.85$\,dB (sphere, \texttt{pavement}).

\paragraph{Comparison}
The same table places AF-NTC, at $25\%$ of the BC7 mip-chain budget, between the two BC7 baselines.
AF-NTC beats trilinear-BC7 on all ten material--scene rows, by $+1.31$ to $+7.63$\,dB, including the highest-frequency \texttt{stone\_wall}.
It stays below aniso-BC7 on every row, by $0.03$ to $3.37$\,dB (mean $1.57$): closest on \texttt{stone\_floor} and furthest on \texttt{pavement} plane.
Renders and error maps for all ten rows are in Appendix~\ref{sec:supp-pairs} (Fig.~\ref{fig:supp-all}).

\begin{table}[t]
\centering
\caption{
    AF-NTC, its ablation, and the two BC7 baselines (PSNR, dB).
    \emph{iso} and \emph{+aniso} are the same decoder trained on isotropic and on anisotropically filtered ground truth; AF-NTC adds LOD conditioning.
    \emph{aniso} and \emph{tri} are aniso-BC7 and trilinear-BC7, which store the full mip chain at $4\times$ the AF-NTC size.
    Materials run from low to high content frequency.
}
\label{tab:main}
\Description{
    A table of PSNR in decibels for five materials, each rendered in a plane scene and a sphere scene, giving ten rows.
    The first three numeric columns are the AF-NTC ablation: an isotropic baseline, the same decoder trained on anisotropic ground truth, and the full AF-NTC with LOD conditioning.
    The last two columns are the BC7 baselines, aniso-BC7 and trilinear-BC7, which store four times more data.
    PSNR rises from the isotropic baseline to plus-aniso to AF-NTC in each of the ten rows.
    AF-NTC is above trilinear-BC7 in every row and below aniso-BC7 in every row.
    Materials are ordered from low to high content frequency.
}
{\footnotesize
\begin{tabular}{@{}llccccc@{}}
\toprule
& & \multicolumn{3}{c}{ours (ablation)} & \multicolumn{2}{c}{BC7, $4\times$ size} \\
\cmidrule(lr){3-5}\cmidrule(lr){6-7}
Material & Scene & iso & +aniso & AF-NTC & aniso & tri \\
\midrule
\texttt{blue\_metal}  & plane  & 50.54 & 51.27 & 51.56 & 53.62 & 45.87 \\
                      & sphere & 50.12 & 51.63 & 51.93 & 53.01 & 47.45 \\
\texttt{clay\_roof}   & plane  & 33.65 & 34.67 & 35.02 & 35.57 & 27.39 \\
                      & sphere & 37.63 & 39.33 & 39.76 & 41.42 & 35.80 \\
\texttt{stone\_floor} & plane  & 33.94 & 36.71 & 38.60 & 39.08 & 35.56 \\
                      & sphere & 34.01 & 37.36 & 39.08 & 39.11 & 34.83 \\
\texttt{pavement}     & plane  & 33.64 & 36.27 & 37.47 & 40.84 & 36.11 \\
                      & sphere & 33.81 & 37.54 & 39.66 & 41.78 & 37.70 \\
\texttt{stone\_wall}  & plane  & 30.27 & 32.68 & 33.48 & 35.09 & 30.21 \\
                      & sphere & 30.18 & 33.45 & 35.48 & 38.24 & 34.17 \\
\bottomrule
\end{tabular}}
\end{table}

\begin{figure}[t]
\centering
\setlength{\tabcolsep}{1pt}
\begin{tabular}{@{}cccc@{}}
    {\scriptsize AF-NTC render} & {\scriptsize trilinear-BC7} & {\scriptsize AF-NTC} & {\scriptsize aniso-BC7} \\
    \includegraphics[width=0.235\columnwidth]{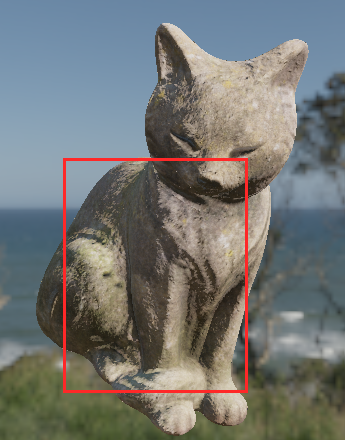}   &
    \includegraphics[width=0.235\columnwidth]{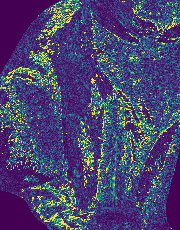}   &
    \includegraphics[width=0.235\columnwidth]{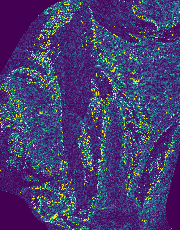}   &
    \includegraphics[width=0.235\columnwidth]{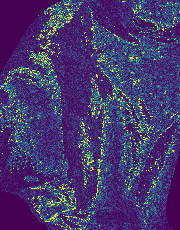} \\
    & {\scriptsize 38.33\,dB} & {\scriptsize 40.34\,dB} & {\scriptsize 40.40\,dB} \\
\end{tabular}
\caption{
    \texttt{cat} mesh.
    \emph{Left}: the AF-NTC render; the red box marks the inset.
    \emph{Right}: $\times16$ absolute error against the supersampled reference over that inset, with whole-frame PSNR.
    Trilinear-BC7 leaves clearly more error, while AF-NTC and the $4\times$-larger aniso-BC7 are within $0.1$\,dB.
}
\Description{
    Four panels for the cat mesh. The leftmost panel is the full AF-NTC render of a concrete cat statue, with a red box marking an inset region on the receding flank.
    The other three panels show the same inset as an absolute error image against the supersampled reference, amplified 16 times and on one shared color scale, for trilinear-BC7, AF-NTC, and aniso-BC7 in that order.
    Their whole-frame PSNR values are 38.33, 40.34, and 40.40 decibels.
    The trilinear-BC7 error panel is visibly brighter; that is, it has more error, while the AF-NTC and aniso-BC7 panels look alike.
}
\label{fig:mesh}
\end{figure}

\paragraph{Meshes}
To test AF-NTC on real 3D models, not just the flat and curved test surfaces, we train it per asset on two PolyHaven meshes whose 4K maps are non-stationary, with a unique UV layout rather than a tiling pattern: a concrete \texttt{cat} statue and a wine \texttt{barrel}.
Table~\ref{tab:mesh} shows PSNR in rendering space on each asset, and Fig.~\ref{fig:mesh} shows the \texttt{cat} images.
At $25\%$ of the BC7 size, AF-NTC stays just below the full-4K aniso-BC7 reference by at most $0.07$\,dB on the two assets and beats the no-AF trilinear-BC7 by $+2.0$ (\texttt{cat}) to $+3.5$\,dB (\texttt{barrel}).
Against the equal-byte, half-resolution (2K) aniso-BC7, AF-NTC is $+0.28$\,dB on \texttt{cat} and $-0.07$\,dB on \texttt{barrel}.
On both assets, the footprints are large (LOD ${\approx}1.4$--$2.4$), so AF already removes most fine detail and a 2K map loses little; we do not claim this equal-byte result for the finer footprints of Table~\ref{tab:main}, where the $2$K chain is $0.4$ to $3.3$\,dB better than AF-NTC.
The \texttt{barrel} images are in Appendix~\ref{sec:supp-mesh} (Fig.~\ref{fig:supp-barrel}).

\begin{table}[t]
\centering
\caption{AF-NTC on mesh assets (PSNR, dB). \emph{aniso} and \emph{tri} are the full-4K BC7 references. \emph{aniso-2K} is the equal-byte 2K aniso-BC7.}
\label{tab:mesh}
\Description{
    A table of PSNR in decibels for two mesh assets, a concrete cat statue and a wine barrel.
    The columns are AF-NTC, the full-resolution 4K aniso-BC7 and trilinear-BC7 references, and an equal-byte half-resolution 2K aniso-BC7.
    AF-NTC is within 0.07 decibels of the full-resolution aniso-BC7 on both assets, is about 2 decibels above trilinear-BC7 on the cat and about 3.5 decibels above it on the barrel, and is within 0.3 decibels of the equal-byte 2K aniso-BC7 on both.
}
{\footnotesize
\begin{tabular}{@{}lcccc@{}}
\toprule
Asset & AF-NTC & aniso & tri & aniso-2K \\
\midrule
\texttt{cat}    & 40.34 & 40.40 & 38.33 & 40.06 \\
\texttt{barrel} & 46.02 & 46.09 & 42.54 & 46.09 \\
\bottomrule
\end{tabular}}
\end{table}

\paragraph{Ablations}
Two ablations on \texttt{pavement} and \texttt{stone\_floor} show where the error comes from; the table and discussion are in Appendix~\ref{sec:supp-ablate} (Table~\ref{tab:ablate}).
Removing the minor-axis level blend costs $5.3$ to $9.3$\,dB, most of it near the camera where footprints are smallest, so the blend is the most important part of the filter.
A $\sim$4$\times$ larger model gains at most $0.4$\,dB: at this budget, the error left is set by compression, not by model size.

\section{Conclusion and Discussion}
\label{sec:conclusion}

We showed that a feature-pyramid NTC decoder can filter anisotropically when given explicit control: a bounded $(\texttt{rot},\texttt{stretch}, \textrm{LOD})$ triple plus a two-axis filter---a tap line along the major axis and a level blend across the minor axis.

Unlike stochastic filtering~\cite{stf2024}, which keeps one decode per pixel but trades aliasing for noise that a temporal post-process must resolve, our filter is deterministic.
A single decode returns a noise-free filtered value that needs no post-processing and is therefore easy to integrate into an existing pipeline, at the cost of more taps per decode.

\paragraph{Scope and limitations}
We claim correct filtering at a fixed compression budget, not a general size--quality win, and we do not claim to match BC7.
AF-NTC comes within $0.5$\,dB of the $4\times$-larger aniso-BC7 on \texttt{stone\_floor}, but stays below it on every row, by up to $3.4$\,dB (Table~\ref{tab:main}).
We report no runtime evaluation: our AF is a PyTorch software emulation, not an optimised GPU implementation on a texture sampler.
How that tap budget trades quality against decoding performance, measured on real hardware, is future work.

\bibliographystyle{ACM-Reference-Format}
\bibliography{ref}

\appendix
\section{Combined Ablations}
\label{sec:supp-ablate}

All numbers in this appendix are PSNR in rendering space, as in Sec.~\ref{sec:results}; higher is better.
Every AF-NTC number is an average over several training runs that differ only in their random seed.

Table~\ref{tab:ablate} gives the filtering and model-size ablations.
Its \emph{AF-NTC} column is our default model, so it repeats the AF-NTC numbers of Table~\ref{tab:main}.

\begin{table}[htb]
\centering
\caption{Combined ablations (dB).
\emph{no-blend} removes the minor-axis level blend from AF-NTC;
\emph{2$\times$ feat.} uses $24$d per-level features (vs. $12$d), and \emph{2$\times$ res.} uses higher-resolution grids ($\sim$4$\times$ params).}
\label{tab:ablate}
\Description{
    A table of PSNR in decibels for two materials, pavement and stone floor, each rendered in a plane scene and a sphere scene, giving four rows.
    The columns are no-blend, which removes the minor-axis level blend from AF-NTC;
    AF-NTC, the default model; two times feature, which doubles the per-level feature dimension; and two times resolution, which uses higher-resolution grids with about four times the parameters.
    Removing the blend costs five to nine decibels.
    Doubling the feature dimension or the grid resolution changes PSNR by at most zero point four decibels in either direction.
}
{\footnotesize\renewcommand{\arraystretch}{0.9}
\begin{tabular}{@{}llcccc@{}}
\toprule
Material & Scene & no-blend & AF-NTC & 2$\times$ feat. & 2$\times$ res. \\
\midrule
\texttt{pavement}     & plane  & 32.16 & 37.47 & 37.45 & 37.63 \\
                      & sphere & 30.36 & 39.66 & 39.67 & 39.83 \\
\texttt{stone\_floor} & plane  & 31.91 & 38.60 & 38.52 & 38.93 \\
                      & sphere & 30.13 & 39.08 & 39.18 & 39.29 \\
\bottomrule
\end{tabular}}
\end{table}

\begin{figure}[htb]
\centering
\setlength{\tabcolsep}{1.5pt}\renewcommand{\arraystretch}{0.8}
\begin{tabular}{@{}cc@{}}
    {\scriptsize no-blend} & {\scriptsize AF-NTC} \\
    \includegraphics[width=0.48\columnwidth]{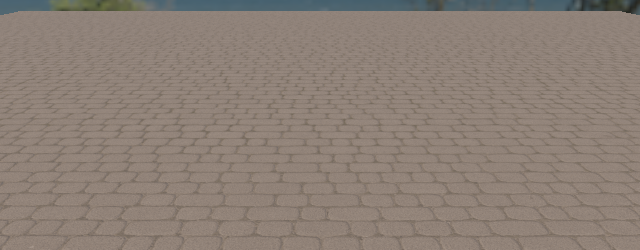} &
    \includegraphics[width=0.48\columnwidth]{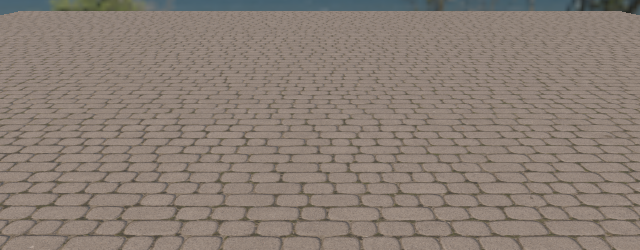} \\[1pt]
    \includegraphics[width=0.48\columnwidth]{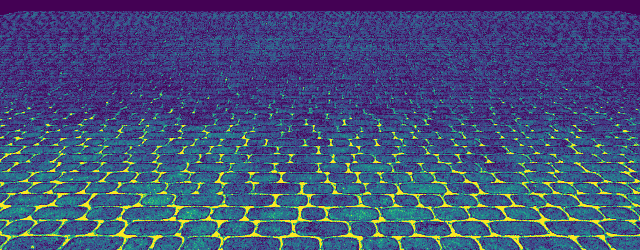} &
    \includegraphics[width=0.48\columnwidth]{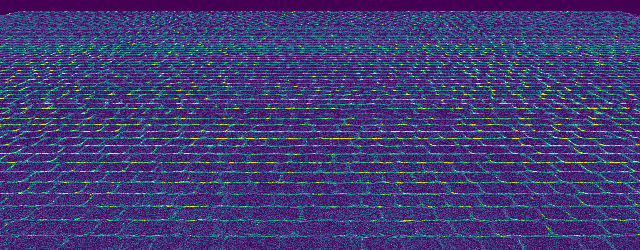} \\
\end{tabular}
\caption{Effect of the minor-axis level blend (\texttt{pavement} plane).
\emph{Top}: rendered crop. \emph{Bottom}: $\times$16 absolute error against the supersampled reference.
Without the blend, the decoder cannot read the grids at a fine enough level for the small footprints near the camera, so the pattern there is washed out.
Further away, the two are almost the same.}
\Description{
    Two columns, no-blend and AF-NTC, for the pavement plane scene.
    The top row shows a rendered crop from each method; the bottom row shows the absolute error against the supersampled reference, amplified sixteen times, on the same crop.
    Without the minor-axis level blend, the near part of the crop, where the footprint is small, loses its surface pattern and the error there is clearly brighter.
    Further from the camera, where the footprint is larger, the no-blend and AF-NTC results look almost the same.
}
\label{fig:supp-blend}
\end{figure}

\paragraph{Filtering}
Our filter has two parts: a tap line along the major axis and a level blend across the minor axis.
The \textit{no-blend} column keeps the tap line and removes the blend, so it still filters along the major axis but no longer across it.
The blend is the more important of the two: without it, PSNR drops by $5.3$ to $9.3$\,dB.
Fig.~\ref{fig:supp-blend} shows where that error is.
Near the camera, the footprints are small, and the decoder has to read the grids at a fine level; without the blend, it cannot, and the pattern is washed out.
Far from the camera, where a coarse level is the right one anyway, the two look the same.

\paragraph{Model size}
We made the model larger in two ways: doubling the feature dimension and using higher-resolution grids.
Both change PSNR by at most $0.4$\,dB (Table~\ref{tab:ablate}).
The error left at this budget is thus set by compression, not by model size: a larger model cannot bring back detail that the footprint filter has already averaged away.

\section{Feeding $(\texttt{rot},\texttt{stretch})$ to the MLP}
\label{sec:supp-mlpinput}

Our decoder gives the multilayer perceptron (MLP) only the level of detail (LOD).
$\texttt{rot}$ and $\texttt{stretch}$ reach it another way: they place the taps, as their orientation and the length of the tap line.
Table~\ref{tab:supp-pe} asks whether the MLP should be given them as an additional $18$d positional encoding.
It should not: the two differ by at most $0.05$\,dB, so we keep the smaller input.

\begin{table}[htb]
\centering
\caption{Giving $(\texttt{rot},\texttt{stretch})$ to the MLP, on \texttt{pavement}.
Ours takes a $49$d input, the $48$d grid features plus LOD;
the $67$d variant adds an $18$d positional encoding of $(\texttt{rot},\texttt{stretch})$.}
\label{tab:supp-pe}
\Description{
    A table of PSNR in decibels for the pavement material in a plane scene and a sphere scene, comparing two decoder inputs.
    The forty-nine-dimensional input is the default: forty-eight grid features plus level of detail.
    The sixty-seven dimensional input adds an eighteen dimensional positional encoding of rotation and stretch.
    The two inputs differ by at most zero point zero five decibels in either scene, so the larger input is not worth its extra size.
}
{\footnotesize\renewcommand{\arraystretch}{0.9}
\begin{tabular}{@{}lccc@{}}
\toprule
Scene & $49$d (ours) & $67$d ($+$PE) \\
\midrule
plane  & 37.53 & 37.48 \\
sphere & 39.74 & 39.71 \\
\bottomrule
\end{tabular}}
\end{table}

\section{Mesh Assets}
\label{sec:supp-mesh}

\begin{figure}[htb]
\centering
\setlength{\tabcolsep}{1.5pt}\renewcommand{\arraystretch}{0.8}
\begin{tabular}{@{}cccc@{}}
    {\scriptsize AF-NTC render} & {\scriptsize trilinear-BC7} & {\scriptsize aniso-BC7} & {\scriptsize AF-NTC} \\
    \includegraphics[width=0.235\columnwidth]{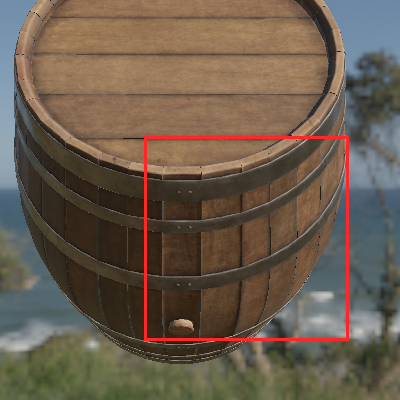}    &
    \includegraphics[width=0.235\columnwidth]{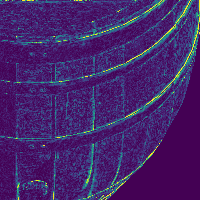}   &
    \includegraphics[width=0.235\columnwidth]{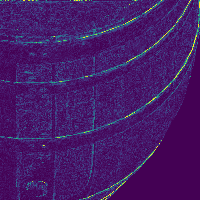} &
    \includegraphics[width=0.235\columnwidth]{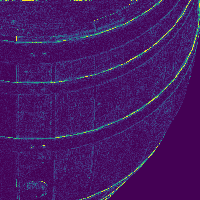}   \\
    & {\scriptsize 42.54\,dB} & {\scriptsize 46.09\,dB} & {\scriptsize 46.02\,dB} \\
\end{tabular}
\caption{\texttt{barrel} mesh.
\emph{Left}: AF-NTC render, with the inset region marked in red.
\emph{Right}: $\times 16$ absolute error against the supersampled reference for each method over that inset;
numbers are whole-frame PSNR.
Over this inset, AF-NTC and aniso-BC7 are within $0.1$\,dB of each other.}
\Description{
    Four panels for the barrel mesh.
    The leftmost panel is the full AF-NTC render of a wine barrel, with a red box marking an inset region on the side of the barrel, where the surface turns away from the camera, and the footprints are the most stretched.
    The other three panels show the same inset as an absolute error image against the supersampled reference, amplified sixteen times, for trilinear-BC7, aniso-BC7, and AF-NTC in that order.
    Their whole-frame PSNR values are 42.54, 46.09, and 46.02 decibels.
    The trilinear-BC7 error panel is visibly brighter on the wooden planks and the metal rings, while the aniso-BC7 and AF-NTC panels look alike.
}
\label{fig:supp-barrel}
\end{figure}

Fig.~\ref{fig:supp-barrel} shows the \texttt{barrel} asset in Table~\ref{tab:mesh}.
The inset covers the side of the barrel, where the surface turns away from the camera, and the footprints are the most stretched.
There, trilinear-BC7 leaves clearly more error on the wooden planks and the metal rings, while AF-NTC and aniso-BC7 look the same.

\newcommand{\galrow}[6]{%
    \multirow{2}{*}{\rotatebox[origin=c]{90}{\scriptsize #3}} &
    \includegraphics[width=\galw]{fig/supp/gal_#1_#2_ref.png}    &
    \includegraphics[width=\galw]{fig/supp/gal_#1_#2_tri.png}    &
    \includegraphics[width=\galw]{fig/supp/gal_#1_#2_aniso.png}  &
    \includegraphics[width=\galw]{fig/supp/gal_#1_#2_ntc.png}    \\
    & &
    \includegraphics[width=\galw]{fig/supp_q/galE_#1_#2_tri.png}   &
    \includegraphics[width=\galw]{fig/supp_q/galE_#1_#2_aniso.png} &
    \includegraphics[width=\galw]{fig/supp_q/galE_#1_#2_ntc.png}   \\
    & & {\scriptsize #4\,dB} & {\scriptsize #5\,dB} & {\scriptsize #6\,dB} \\
}

\newcommand{\galhead}[1]{%
    \multicolumn{5}{@{}c@{}}{\scriptsize\texttt{#1} scene} \\
    & {\scriptsize Reference} & {\scriptsize trilinear-BC7} & {\scriptsize aniso-BC7} & {\scriptsize AF-NTC} \\
}

\begin{figure*}[p]
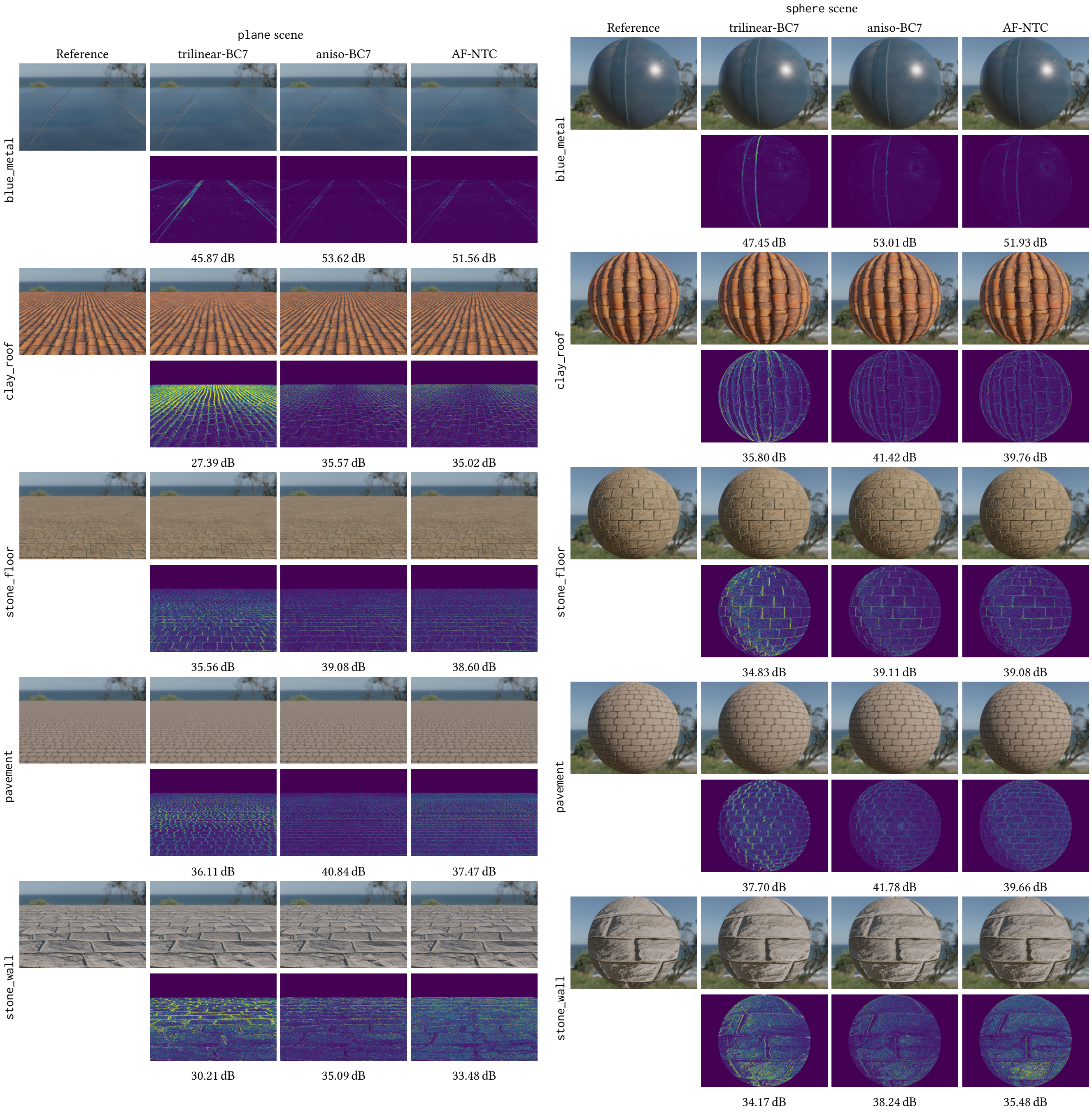

\centering
\setlength{\tabcolsep}{1pt}\renewcommand{\arraystretch}{0.8}
\begin{tabular}{@{}c@{\hspace{8pt}}c@{}}
\begin{tabular}{@{}ccccc@{}}
    \galhead{plane}
    \galrow{plane}{bluemetal}{\texttt{blue\_metal}}{45.87}{53.62}{51.56}
    \galrow{plane}{clayroof}{\texttt{clay\_roof}}{27.39}{35.57}{35.02}
    \galrow{plane}{stonefloor}{\texttt{stone\_floor}}{35.56}{39.08}{38.60}
    \galrow{plane}{pavement}{\texttt{pavement}}{36.11}{40.84}{37.47}
    \galrow{plane}{stonewall}{\texttt{stone\_wall}}{30.21}{35.09}{33.48}
\end{tabular}
&
\begin{tabular}{@{}ccccc@{}}
    \galhead{sphere}
    \galrow{sphere}{bluemetal}{\texttt{blue\_metal}}{47.45}{53.01}{51.93}
    \galrow{sphere}{clayroof}{\texttt{clay\_roof}}{35.80}{41.42}{39.76}
    \galrow{sphere}{stonefloor}{\texttt{stone\_floor}}{34.83}{39.11}{39.08}
    \galrow{sphere}{pavement}{\texttt{pavement}}{37.70}{41.78}{39.66}
    \galrow{sphere}{stonewall}{\texttt{stone\_wall}}{34.17}{38.24}{35.48}
\end{tabular}
\end{tabular}
\caption{All ten material--scene pairs: the \texttt{plane} scene on the left, the \texttt{sphere} scene on the right, five materials each, ordered from low to high content frequency.
Each material has a row of renders and the $\times16$ absolute error against the supersampled reference for the three compressed methods.
Every material scores higher on the sphere than on the plane.}
\Description{
    A grid of renders for all ten material-scene pairs, the plane scene on the left and the sphere scene on the right, five materials each:
    blue metal, clay roof, stone floor, pavement and stone wall, ordered from low to high content frequency.
    Each material has a row with a reference render, a trilinear-BC7 render, an aniso-BC7 render, an AF-NTC render, and the absolute error of the last three against the supersampled reference, amplified sixteen times, with whole-frame PSNR below each error image.
    Across every row, trilinear-BC7 loses the surface pattern wherever the footprint is stretched, most visibly on the clay roof plane, while AF-NTC keeps the pattern and is visually close to the four times larger aniso-BC7.
    Every material scores higher in the sphere scene than in the plane scene.
}
\label{fig:supp-all}
\end{figure*}

\section{All Ten Material--Scene Pairs}
\label{sec:supp-pairs}

Sec.~\ref{sec:results} shows only one of the ten material--scene rows of Table~\ref{tab:main} as a picture, the \texttt{pavement} plane in the teaser (Fig.~\ref{fig:teaser}); Fig.~\ref{fig:supp-all} shows all ten.
Trilinear-BC7 blurs the surface pattern away wherever the footprint is stretched, most visibly on the \texttt{clay\_roof} plane, the row where AF-NTC gains the most over it ($+7.6$\,dB).
AF-NTC keeps the pattern, and on most rows its error is close to that of the $4\times$-larger aniso-BC7; it is worse only on the \texttt{pavement} plane ($-3.4$\,dB) and the \texttt{stone\_wall} sphere ($-2.8$\,dB), and even there the two stay visually close.

\end{document}